\documentclass[prl,showpacs,superscriptaddress,twocolumn]{revtex4-1}

\usepackage{graphicx, subfigure}

\usepackage{amsmath}
\usepackage{amsfonts}
\usepackage{graphicx}

\usepackage{natbib}
\usepackage[final=true]{hyperref}
\hypersetup{
    colorlinks=true,
    linkcolor=Maroon,
    filecolor=magenta,      
    urlcolor=cyan,
    citecolor=PineGreen,    
}

\usepackage{color}
\usepackage[usenames,dvipsnames]{xcolor}

\usepackage{xcolor}
\usepackage[dvipsnames]{xcolor}

\newcommand{\X}{_{\mathrm{X}}}
\newcommand{\ph}{_{\mathrm{Ph}}}
\newcommand{\rt}{({\bf r},t)}

\begin{document}

\title{Parallel dark soliton pair in a bistable 2D exciton-polariton superfluid}

\author{G.~Lerario}
\email{giovanni.lerario@lkb.upmc.fr}
\affiliation{Laboratoire Kastler Brossel, Sorbonne Universit\'e, CNRS,ENS-PSL Research University, Coll\`ege de France, 4 place Jussieu, 75252 Paris, France}
\affiliation{CNR NANOTEC, Istituto di Nanotecnologia, via Monteroni, 73100 Lecce, Italy}

\author{S.~V.~Koniakhin}
\affiliation{Institut Pascal, PHOTON-N2, Universit\'e Clermont Auvergne, CNRS, SIGMA Clermont, Institut Pascal, F-63000 Clermont-Ferrand, France}
\affiliation{St. Petersburg Academic University - Nanotechnology Research and Education Centre of the Russian Academy of Sciences, 194021 St. Petersburg, Russia}

\author{A. Ma\^{i}tre}
\affiliation{Laboratoire Kastler Brossel, Sorbonne Universit\'e, CNRS,ENS-PSL Research University, Coll\`ege de France, 4 place Jussieu, 75252 Paris, France}

\author{D.~Solnyshkov}
\affiliation{Institut Pascal, PHOTON-N2, Universit\'e Clermont Auvergne, CNRS, SIGMA Clermont, Institut Pascal, F-63000 Clermont-Ferrand, France}
\affiliation{Institut Universitaire de France (IUF), 1 rue Descartes, 75231 Paris, France}

\author{A. Zilio}
\affiliation{Université de Paris, Laboratoire Jacques-Louis Lions (LJLL), F-75013 Paris, France.}
\affiliation{Sorbonne Université, CNRS, LJLL, F-75005 Paris, FranceUniversité Paris-Diderot, Sorbonne Paris-Cité, Laboratoire Jacques-Louis Lions, CNRS, Bâtiment Sophie Germain, 75205 Paris CEDEX 13, France}

\author{Q. Glorieux}
\affiliation{Laboratoire Kastler Brossel, Sorbonne Universit\'e, CNRS,ENS-PSL Research University, Coll\`ege de France, 4 place Jussieu, 75252 Paris, France}
\affiliation{Institut Universitaire de France (IUF), 1 rue Descartes, 75231 Paris, France}

\author{G.~Malpuech}
\affiliation{Institut Pascal, PHOTON-N2, Universit\'e Clermont Auvergne, CNRS, SIGMA Clermont, Institut Pascal, F-63000 Clermont-Ferrand, France}

\author{E. Giacobino}
\affiliation{Laboratoire Kastler Brossel, Sorbonne Universit\'e, CNRS,ENS-PSL Research University, Coll\`ege de France, 4 place Jussieu, 75252 Paris, France}

\author{S. Pigeon}
\affiliation{Laboratoire Kastler Brossel, Sorbonne Universit\'e, CNRS,ENS-PSL Research University, Coll\`ege de France, 4 place Jussieu, 75252 Paris, France}

\author{A. Bramati}
\email{alberto.bramati@lkb.upmc.fr}
\affiliation{Laboratoire Kastler Brossel, Sorbonne Universit\'e, CNRS,ENS-PSL Research University, Coll\`ege de France, 4 place Jussieu, 75252 Paris, France}
\affiliation{Institut Universitaire de France (IUF), 1 rue Descartes, 75231 Paris, France}

\begin{abstract}
Collective excitations, such as vortex-antivortex and dark solitons, are among the most fascinating effects of macroscopic quantum states. However, 2D dark solitons are unstable and collapse into vortices due to snake instabilities. Making use of the optical bistability in exciton-polariton microcavities, we demonstrate that a pair of dark solitons can be formed in the wake of an obstacle in a polariton flow resonantly supported by a homogeneous laser beam. Unlike the purely dissipative case where the solitons are grey and spatially separate, here the two solitons are fully dark, rapidly align at a specific separation distance and propagate parallel as long as the flow is in the bistable regime. Remarkably, the use of this regime allows to avoid the phase fixing arising in resonant pumping regime and to circumvent the polariton decay. Our work opens very wide perspectives of studying new classes of phase-density defects which can form in driven-dissipative quantum fluids of light.
\end{abstract}

\maketitle

 Promoted by the demonstration of Bose Einstein condensation (BEC) \cite{davis1995bose_REF1,kasprzak2006bose_REF2}, collective excitations have been extensively studied in cold atoms and exciton-polariton systems \cite{anglin2002bose_REF3,byrnes2014exciton_REF4}. During the last two decades, topologically protected vortices \cite{matthews1999vortices_REF5,burger1999dark_REF6,lagoudakis2008quantized_REF7,roumpos2011single_REF8,tosi2012geometrically_REF9,sanvitto2011all_REF10,lerario2014room_REF11,Gauthier2019,Johnstone2019} and dark solitons \cite{burger1999dark_REF6,grosso2011soliton_REF12,amo2011polariton_REF13,Wertz2012} have been successfully demonstrated theoretically and experimentally in both  systems.

 Exciton-polaritons are bosonic quasi-particles resulting from the exciton-photon strong coupling in microcavities \cite{Microcavities}, which gives them hybrid properties coming from their components. They have a very light mass inherited by the photon component and they interact with each other due to the exciton-exciton interaction. In recent years, these systems became very attractive in the context of out-of-equilibrium condensates and 2D quantum fluid hydrodynamics \cite{Carusotto2013}. Exciton-polaritons can be created by optical excitation. In particular, it is possible to create metastable supersonic flows because of the very weak thermal relaxation. This unique possibility allowed the hydrodynamic generation of dark solitons using a supersonic polariton wavepacket hitting a structural defect in its in-plane propagation \cite{grosso2011soliton_REF12,amo2011polariton_REF13}. Both pulsed and continuous-wave (cw) resonant excitation have been used. In pulsed resonant configuration, the polariton flow can propagate freely after the pulse is over, but its lifetime and propagation distance are relatively short. In cw resonant configuration, the pump should not spatially overlap the regions where solitons are created (otherwise the phase is fixed by the laser and phase defects cannot exist). As a result, similarly to the pulsed case, the flow strongly decays along the propagation. 

The formation of oblique dark solitons within the polariton coherent state has been observed in such configuration \cite{amo2011polariton_REF13}. The transverse ''snake'' instability \cite{Kuznetsov1988,PhysRevA.65.043612}, which normally destroys the 2D dark solitons by converting them into chains of vortex-antivortex pairs, is suppressed by the supersonic flow \cite{kamchatnov2008stabilization_REF22}. The 2D solitons are effectively 1D in this case, and the spatial direction along the polariton flow (in-plane wave vector of laser light) is mapped to an effective time. In 1D cavities the stable solitons were also studied and their phase relation with respect to the phase of the laser was extensively investigated \cite{Goblot2016}. 

Importantly, recent theoretical papers have highlighted that topological excitations (vortex-antivortex pairs and bound dark soliton pairs) can be stabilized in driven dissipative condensates when working within the optical bistability regime, overcoming the phase fixing problem \cite{pigeon2017sustained_REF17,koniakhin2019stationary_REF17b,parra2016dark_REF18}. Indeed within the bistable hysteresis cycle, two regions are available for the system, respectively at high and low polariton density.  In the high-density regions above the bistability regime the phase is fixed by the resonant laser. On the contrary, the regions of the bistability regime can have an arbitrary phase, which enables the existence of a rich variety of topological defects, while the radiative decay of the polaritons is compensated, allowing long propagation distances in cw experiment.


In this paper, we report the observation of new quantum hydrodynamic effects revealed by using the optical bistability of exciton-polaritons in microcavities. We experimentally demonstrate the hydrodynamic generation of a parallel dark soliton pair in the hydrodynamic wake of a cavity structural defect. This bound soliton pair propagates along the polariton flow as long as the laser pumping sustains it, thus removing the constraints imposed by the polariton lifetime in previous experimental configurations.

\begin{figure}
\centering
\includegraphics[width=0.49\textwidth]{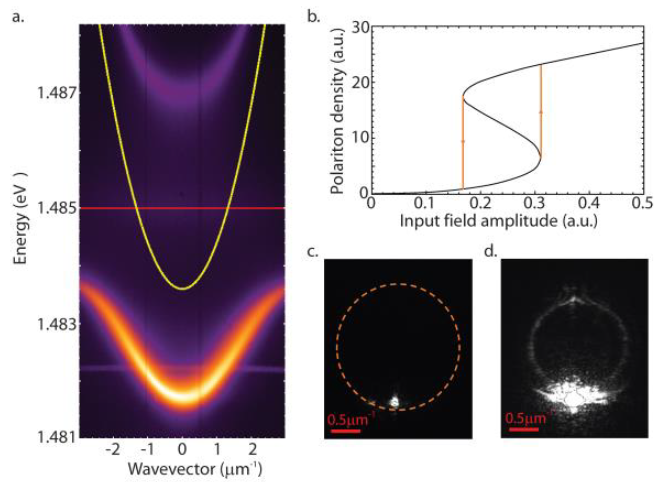}
\caption{\label{fig_1} Cavity dispersion and bistability. a. Energy dispersion of the exciton-polariton emission (with logarithmic scale for the intensity) when the microcavity is pumped with a non-resonant cw laser at 1.54 eV. The bare photon and exciton dispersions are in yellow and red curves, respectively. b. Theoretical calculations of the bistability loop according to the experimental parameters given in the text. c. Output intensity distribution in the momentum space when the system is in the lower branch of the bistability loop and with an excitation at $k= 1.2 \mu$m$^{-1}$ and energy 1.4824 eV. The orange dashed circle depicts the Rayleigh scattering ring, which is visible for longer integration time (higher contrast). d. Output intensity distribution in the momentum space when the system locks to the upper branch of the bistability loop. The scattering ring is now visible and it has a smaller radius compared to the one in panel c.}
\end{figure}

The device under investigation is a GaAs/AlGaAs microcavity with 21/24 (front/back) layers of DBR and In$_{0.04}$Ga$_{0.96}$As quantum wells at each of the three antinodes of the confined electromagnetic field \cite{PhysRevB.61.R13333}. All experiments are performed at 10 K. The exciton energy is 1.485 eV, the cavity exciton-photon detuning is negative (-1.4 meV) and the half Rabi splitting is 2.55~meV (see Fig.~\ref{fig_1}a). The polariton mass, extracted from the dispersion,  is $5\cdot 10^{-5}$ free electron mass. The polariton lifetime is 14 ps. The experiments are performed in transmission configuration, i.e. the excitation and the detection are on opposite sides of the sample. A continuous wave (cw) single mode laser excites the polaritons in the quasi-resonant regime. In order to work within the optical bistability regime, the laser frequency is slightly blue detuned (0.16 meV) with respect to the lower polariton dispersion branch, guaranteeing the generation of a hysteresis loop \cite{baas2004optical_REF19}, see Fig. \ref{fig_1}b.

\begin{figure*}
\centering
\includegraphics[width=0.95\textwidth]{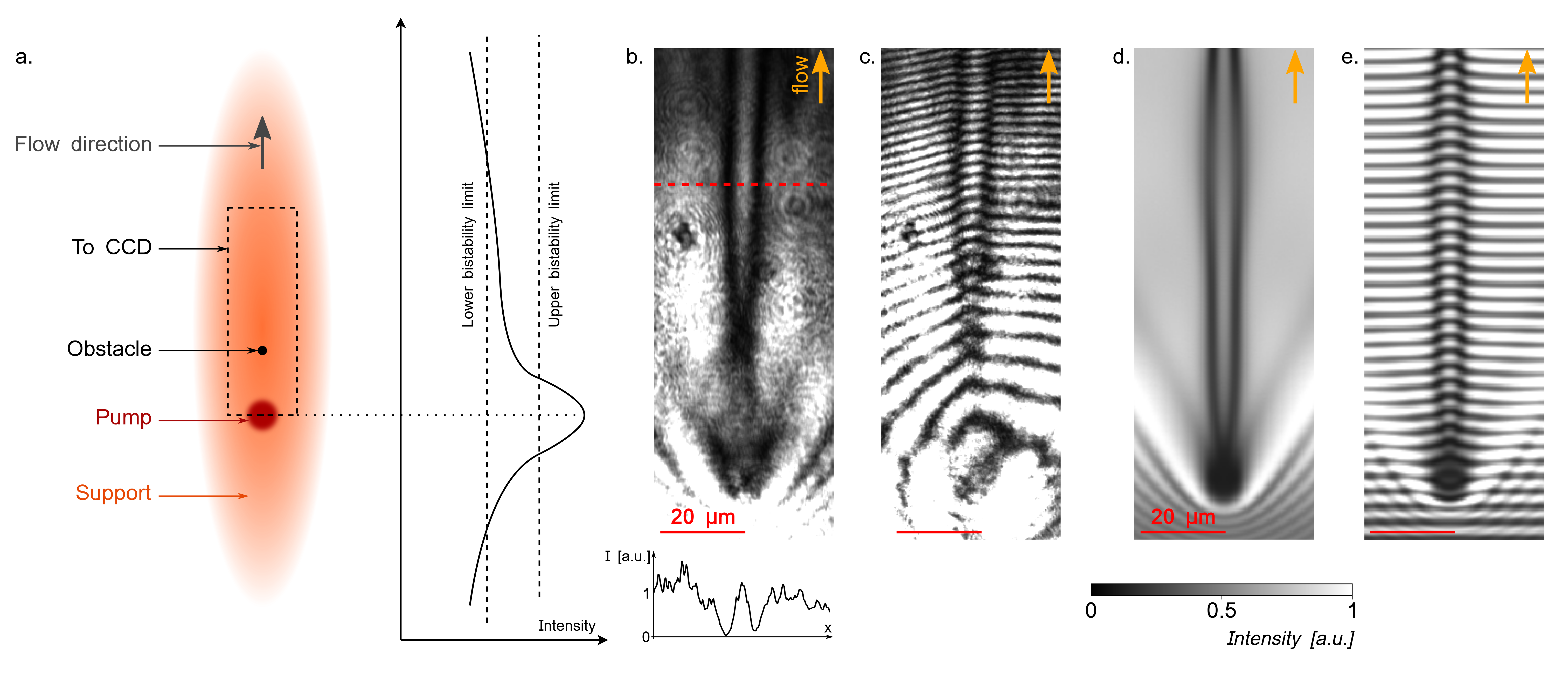}
\caption{\label{fig_2} a. Experiment scheme with pump and support beams. The curve sketches the laser intensity along the propagation axe. b. Intensity map of the polariton fluid flowing across a structural defect. The turbulences generated in the wake of the defect evolve into two parallel dark solitons. The red dotted line indicates the position of the transverse section presented in the inset
c. Interference pattern associated to panel b, showing a phase jump close to $\pi$ across the solitons. d. Theoretical simulations (using the driven dissipative Gross-Pitaevskii equation) of the intensity map for parameters corresponding to those used in the experiment. e. Theoretical interference pattern corresponding to panel d.}
\end{figure*}

The present configuration implements the effective pump/support scheme proposed in \cite{pigeon2017sustained_REF17} for sustaining the propagation of topological defects in dissipative polariton quantum fluid. In such scheme, a strong localized pump drives the system to the upper bistability branch. This pump is surrounded by a weaker support beam with the intensity falling within a bistability loop. This condition allows to overcome the polariton decay while avoiding the phase fixing. In our experiment, the pump beam center is positioned slightly upstream of an obstacle (cavity structural defect) and its power is chosen to efficiently drive the system to the upper branch of bistability loop (Fig.~\ref{fig_2}(a)). Concerning the support beam, cylindrical lenses are used to shape it and to make it elliptic (100x400 $\mu$m FWHM) with the flow direction along the major ellipse axis. The pump beam is near the center of the support beam. As far as both these beams are obtained by splitting the same initial laser beam (elsewhere in the optical path of the setup), they remain mutually coherent.

The polariton group velocity is finely tuned by choosing the angle of incidence $\theta$ of the pump. Indeed, the in-plane wavevector ($k=k_0 \sin \theta \approx  1.2$ $\mu$m$^{-1}$ (see Fig. \ref{fig_1}c and d), where $k_0$ is the wavevector of the pumping laser) determines the group velocity according to the polariton dispersion relation ($v_{flow} = 1.52$ $\mu$m/ps) reported in Fig.~\ref{fig_1}a.  In such quantum fluid, the sound speed reads $c_s=\sqrt{g |\psi|^2/m}$, where $g$ is the polariton-polariton interaction constant, $m$ the polariton mass and $|\psi|^2$ the polariton density. In our experimental conditions $c_s\approx0.4~\mu$m/ps, which means that the obtained flow is supersonic. The healing length is $\xi = \hbar/\sqrt{2m g |\psi|^2} = \hbar/(\sqrt{2}mc_{s}) \approx 2.2~\mu$m.


In the geometry described above, with the support beam close to the lower limit of the upper bistability branch, we observe the hydrodynamic formation of a soliton pair in the shadow of the defect (Fig.~\ref{fig_2}b). The intensity of the pump does not strongly affect the results because its only role is to locally maintain the system to the upper bistability branch. During the first 40 $\mu$m of their propagation, the solitons reach an equilibrium separation distance (about 8 $\mu$m) and then they continue a parallel propagation for about the next 45 $\mu$m. The inset of fig. \ref{fig_2}b represents a transverse section across the solitons corresponding to the red line in the main figure. It clearly shows that the solitons are fully dark with a density depth reaching zero at the center of the soliton. From the interference map obtained by homodyne detection (Fig. \ref{fig_2}c), one also sees that the phase jump across each soliton is very close to $\pi$ when the two dark solitons align to each other, which means their transverse velocity is close to zero.


To reproduce the soliton behavior observed in the experiments, we have performed numerical simulations based on the coupled equations for the excitons ($\psi_X$) and cavity photon ($\psi_{Ph}$) fields:

\begin{widetext}
\begin{equation*}
i\hbar\frac{\partial \psi\ph\rt }{\partial t}  =
\left[-\frac{\hbar^2\nabla
^2}{2m} + V({\bf r}) - i\Gamma\ph \right]\psi\ph\rt + V \psi\X\rt + (S({\bf r})+P({\bf r}))e^{-i \omega_0 t},
\label{eq_Ph}
\end{equation*}

\begin{equation*}
i\hbar\frac{\partial \psi\X\rt }{\partial t}  =
\left[ V({\bf r}) + g_X\vert\psi\X\rt \vert^2 - i\Gamma\X
- \Delta\X \right]\psi\X\rt + V \psi\ph\rt.
\label{eq_X}
\end{equation*}
\end{widetext}

The parameters (cavity photon mass $m$, laser energy $\omega_0$, photon lifetimes $\Gamma\ph$, half-Rabi splitting $V$ and cavity-exciton detuning $\Delta\X$) were taken to be the same as in the experiment. The exciton lifetime $\Gamma\X$ was taken to be 150~ps.  The obstacle was modelled as a 10 meV potential barrier $V({\bf r})$ with a Gaussian shape of 10 $\mu$m width. The relative positions of the obstacle and pump beam also reproduce the experimental configuration. $P({\bf r})$ and $S({\bf r})$ describe the spatial profiles of the pump and support beams, respectively, and their magnitudes are adjustable parameters. The output images reflect the spatial profile of the photon component density $|\psi\ph\rt|^2$ and the interference pattern $|\psi\ph\rt + A_0 e^{-i\mathbf{k_0r}}|^2$ . The results of the simulations (Figures~\ref{fig_2}d and e) are in excellent agreement with the corresponding experimental images.

The homogeneous high density areas located on both sides of the soliton pair have the same phase as the support beam, due to the resonant pumping. On the other hand, dark solitons are characterized by a zero polariton density inside the soliton and a phase jumps of $\pi$ across the soliton. 
The propagation distance of the soliton pair is limited by both the sample wedge and the finite size of the support laser. At some distance, due to the Gaussian shape of the support beam, the flow falls in the linear regime, with low polariton density and the soliton pair can not be sustained anymore.

The existence of an equilibrium distance implies a minimum in the potential energy. As well known in the literature \cite{kivshar1995lagrangian_REF20}, there is a repulsive potential between the two solitons in an undriven system, which increases when the solitons separation distance decreases, so they cannot merge with each other for small value of the solitons transverse speed. 
On the other hand, the expansion of the area in between the solitons is inhibited by the presence of the support beam which is out of phase (i.e. the forcing term in the GPE results in an increase of the potential energy at increasing solitons distance). These two opposite contributions stabilize the solitonic structure similarly to the role of the trapping potential in atomic condensates\cite{kamchatnov2008stabilization_REF22,muryshev1999stability_REF23,weller2008experimental_REF24,theocharis2010multiple_REF25}. As a consequence, the total potential is diatomic-like, it traps the dark solitons in proximity to a specific equilibrium separation distance allowing the generation of the bound and stable soliton pair.


\begin{figure}
\centering
\includegraphics[width=0.45\textwidth]{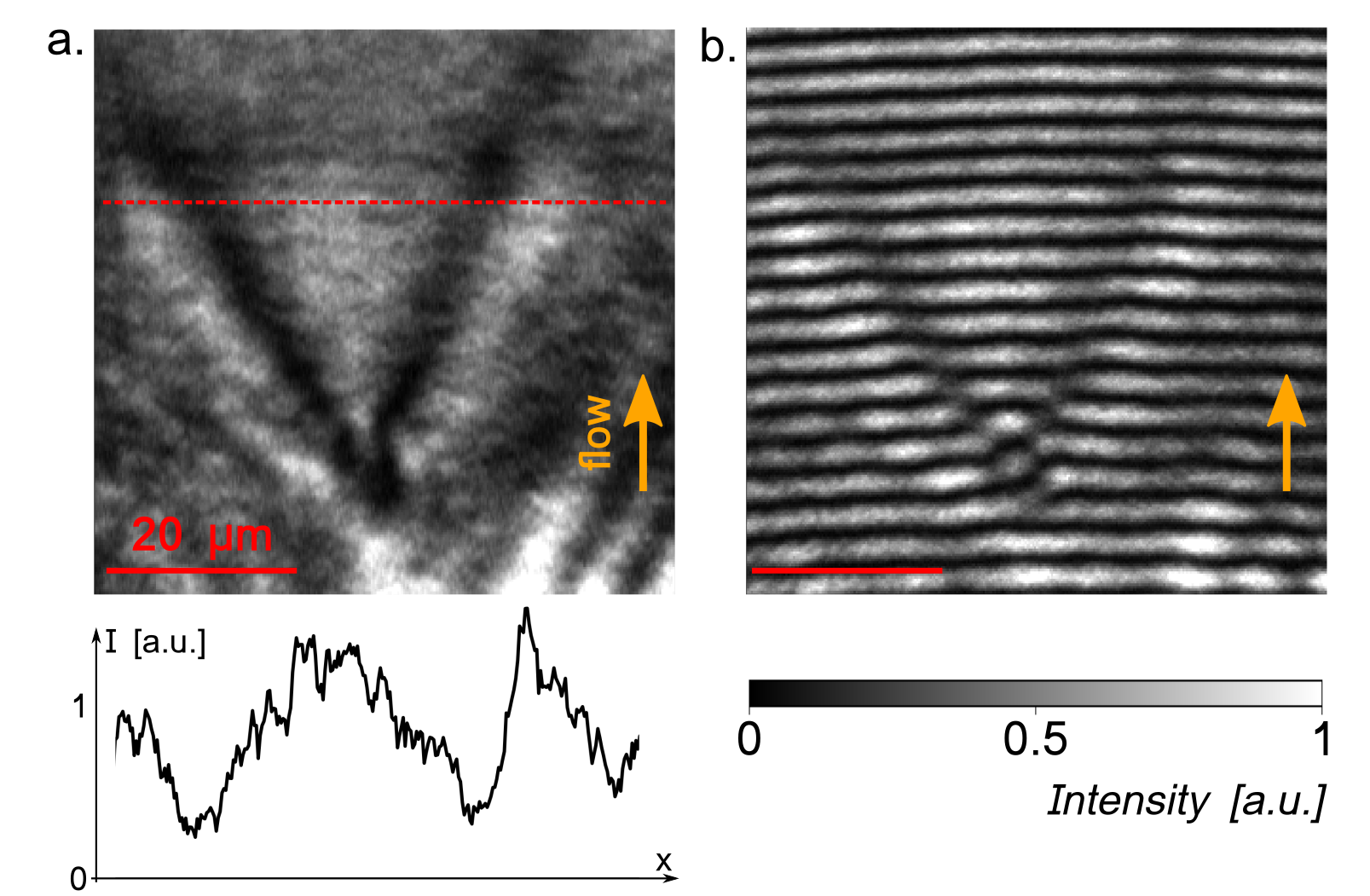}
\caption{\label{fig_3} Oblique solitons. a. Intensity map of the polariton flowing across a structural defect and excited using a semi-circular shaped pump upstream to the defect. The generated solitons in the wake of the defect are grey, as shown in the inset corresponding to the transverse section indicated by the red dotted line, and they propagate with increasing separation distance. b. Interference pattern associated to panel a. The phase jump is decreasing along the propagation as the solitons vanish.}
\end{figure}

In order to compare the bound soliton pair to the solitons generated in a purely dissipative regime, we have performed an experiment similar to the one in Ref.~\cite{amo2011polariton_REF13}, that is, removing the support beam(Fig. \ref{fig_3}). 
In order to avoid the phase fixing of the polaritons to the laser phase, which inhibits the solitons generation, the pump beam is exiting the sample at a distance of about 20 $\mu$m far from the defect position, thus it does not overlap with the structural defect. The solitons, spontaneously generated along the flow of the fluid, are oblique. In this case, differently from the previous configuration, the expansion of the inner region between the solitons is not inhibited by the presence of the out-of-phase support beam. Therefore, the separation distance between the solitons increases along the flow, due to their repulsive interaction. Moreover, (see the inset of figure 3) the oblique solitons are grey and  they have a large width (10 $\mu$m FWHM) compared to the width observed for parallel solitons obtained in the driven-dissipative regime with the support beam and within the bistability loop (2.8 $\mu$m FWHM). 

Furthermore, the oblique solitons vanish soon because of the decrease of the mean polariton density without the support. Indeed, the fluid propagation length is 19.6 $\mu$m ($v_{flow}$=1.4 $\mu$m/ps) and it limits the soliton propagation distance at about 30 $\mu$m. 

\begin{figure}[h]
\centering
\includegraphics[width=0.45\textwidth]{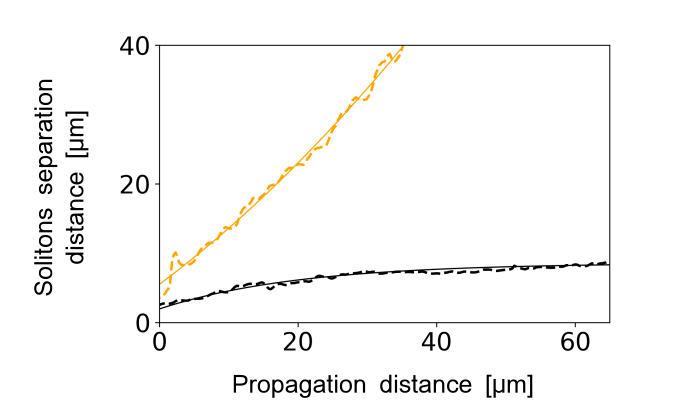}
\caption{\label{fig_4} Solitons separation distance. Dark solitons separation distance for parallel (black dashed line) and oblique (orange dashed line) solitons along their in-plane propagation. Black and orange solid lines are the fitting of the experimental data (see main text for further details).}
\end{figure}

Fig. \ref{fig_4} shows the separation distance of the two kinds of solitons, oblique and parallel, along their propagation (orange and black dashed lines respectively). The black solid line in Fig.~\ref{fig_4} is the fit of the experimental soliton distance with the function $2d_0(1-\exp(-x/d_l))$. The initial transient region, where the solitons are oblique, has a characteristic length $d_l=14$ $\mu$m, then the solitons reach the equilibrium distance, with an asymptotic separation of $2d_0=8$ $\mu$m. On the other hand, the separation distance of the oblique solitons from Fig.~\ref{fig_3} keeps increasing with the propagation, as shown by its polynomial fit (orange solid line of Fig.\ref{fig_4}).

Our experimental results demonstrate that collective excitations with rich phase behavior can be generated in a polariton superfluid within the bistable regime and detected on large length scales in cw experiments. The presence of the support beam allows the formation of a stable bound soliton pair. The key advantage of this experimental configuration is the possibility to decouple the collective excitations lifetime from the polariton one. Indeed, while the solitons propagation length without support is limited by the polariton lifetime, in the bistable regime the collective excitations are sustained by a constant polariton density along their propagation and, therefore, they can propagate as long as the system is on the bistability regime. In our experiment the bound solitons propagate on a much longer distance (about 80 $\mu$m) than the obliques solitons without support beam (about 30 $\mu$m).  These results open new perspectives for probing the hydrodynamic evolution of collective excitations, quantum turbulence and mesoscopic-scale physics in out-of-equilibrium condensates.


\begin{acknowledgments}
We acknowledge the support of the ANR projects "Quantum Fluids of Light" (ANR-16-CE30-0021) and C-FLigHT (ANR-16-ACHN-0027) and of the ANR program "Investissements d'Avenir" through the IDEX-ISITE initiative 16-IDEX-0001 (CAP 20-25). This work has received funding from the European Union’s Horizon 2020 research and innovation programme under grant agreement No. 820392 (PhoQuS). S.V.K. acknowledges the support from the Ministry of Education and Science of Russian Federation (Project 16.9790.2019). G.L. would like to thank Iacopo Carusotto for useful discussions.
\end{acknowledgments}

\bibliography{bib}

\end{document}